\documentclass[twocolumn]{aastex63}
\usepackage{color}
\usepackage{graphicx}
\usepackage{amsmath}
\usepackage{amssymb}	
\usepackage{diagbox}
\usepackage{threeparttable}

\newcommand{\mc}{\mathcal}
\usepackage{ulem}
\usepackage{natbib}

\shorttitle{The radial distribution of pulsars}
\shortauthors{Xie et al.}

\begin{document}

\title{Modelling The Radial Distribution of Pulsars in the Galaxy}
\correspondingauthor{J. B. Wang; Na Wang}
\email{1983wangjingbo@163.com;na.wang@xao.ac.cn}

\author{J. T. Xie}
\affiliation{Zhejiang Lab, Hangzhou, Zhejiang 311121, China\\}
\affiliation{Xinjiang Astronomical Observatory, Chinese Academy of Sciences, 150 Science 1-Street, Urumqi, Xinjiang 830011, China\\}

\author{J. B. Wang}
\affiliation{Institute of Optoelectronic Technology, Lishui University, Lishui, Zhejiang 323000, China\\}

\author{N. Wang}
\affiliation{Xinjiang Astronomical Observatory, Chinese Academy of Sciences, 150 Science 1-Street, Urumqi, Xinjiang 830011, China\\}
\affiliation{Key Laboratory of Radio Astronomy, Chinese Academy of Sciences, 150 Science 1-Street, Urumqi, Xinjiang, 830011, China\\}
\affiliation{Xinjiang Key Laboratory of Radio Astrophysics, 150 Science1-Street, Urumqi, Xinjiang, 830011, China}

\author{R. Manchester}
\affiliation{Australia Telescope National Facility, CSIRO Space and Astronomy, PO Box 76, Epping, NSW 1710, Australia\\}

\author{G. Hobbs}
\affiliation{Australia Telescope National Facility, CSIRO Space and Astronomy, PO Box 76, Epping, NSW 1710, Australia\\}

\begin{abstract}

The Parkes 20 cm Multibeam pulsar surveys have discovered nearly half of the known pulsars and revealed many distant pulsars with high dispersion measures. Using a sample of 1,301 pulsars from these surveys, we have explored the spatial distribution and birth rate of normal pulsars. The pulsar distances used to calculate the pulsar surface density are estimated from the YMW16 electron-density model. 
When estimating the impact of the Galactic background radiation on our survey, we projected pulsars in the Galaxy onto the Galactic plane, assuming that the flux density distribution of pulsars is uniform in all directions, and utilized the most up-to-date background temperature map. We also used an up-to-date version of the ATNF Pulsar Catalogue to model the distribution of pulsar flux densities at 1400 MHz.
We derive an improved radial distribution for the pulsar surface density projected on to the Galactic plane, which has a maximum value at $\sim$4 kpc from the Galactic Centre. We also derive the local surface density and birthrate of pulsars, obtaining 47 $\pm$ 5 $\mathrm{kpc^{-2}}$ and $\sim$ 4.7 $\pm$ 0.5 $\mathrm{kpc^{-2}\ Myr^{-1}}$, respectively.
For the total number of potentially detectable pulsars in the Galaxy, we obtain (1.1 $\pm$ 0.2) $\times$ $10^{4}$ and (1.1 $\pm$ 0.2) $\times$ $10^{5}$ before and after applying the \citet{TM98} beaming correction model. 
The radial distribution function is used to estimate the proportion of pulsars in each spiral arm and the Galactic centre.

\end{abstract}


\section{Introduction} \label{sec:intro}

Neutron stars (NSs) are thought to form in core-collapse supernovae with Population I OB stars as progenitors \citep{1934PNAS...20..254B}. Most of the OB stars are located in the Galactic plane and trace the structure of the Galactic spiral arms \citep{2019MNRAS.487.1400C}. Many NSs are found as radio pulsars in our Galaxy. The observed pulsars also show an association with the spiral arms, and most of them are close to the Galactic plane. 

Many pulsar surveys have been carried out since the discovery of the first pulsar. At the time of writing, more than 3300 radio pulsars have been discovered.\footnote{https://www.atnf.csiro.au/research/pulsar/psrcat/ (V1.70) \citep{2005AJ....129.1993M}} They represent about 10 to 25 percent of the potentially detectable pulsars beaming towards us in the Galaxy \citep[][]{2006ApJ...643..332F,lfl+06}. Models for pulsar the galactocentric radial and height distributions, luminosity distribution, magnetic field evolution and radiation model, etc. have all been established \citep[][]{2004A&A...422..545Y,2006MNRAS.372..777L,2006ApJ...643..332F,2013MNRAS.431.1352B,2013MNRAS.434.1387L}. 
The properties of the radio pulsar population provide valuable constraints on the evolution of NSs. For instance, these results can be used to estimate the birthrate of pulsars and model the Galactic structure. Since the pulsar progenitors, O-B stars, or their associated HII regions, actually define the spiral structure of our Galaxy \citep{2014A&A...569A.125H,2019MNRAS.487.1400C}, it is expected that at least young pulsars will be concentrated in spiral arms. From the Galactic longitude distribution of known pulsars, \citet{kbm+03} and \citet{2006ApJ...643..332F} showed that such concentrations exist. Consequently, it is necessary to model such structures in any realistic pulsar population synthesis simulation.

Pulsars in the Galactic plane (and, in particular, close to the Galactic centre; GC) are relatively hard to detect as the sky background temperature and the interstellar medium (ISM) electron density are higher towards the Galactic plane than in other directions, reducing the survey sensitivity.
These issues are less significant at higher observing frequencies, but pulsar flux densities generally decrease with increasing observing frequency. Early pulsar surveys were mostly carried out at relatively low frequencies.
Pulsar surveys at 1.4 GHz using Parkes telescope (e.g., \citealt{2001MNRAS.328...17M}) revealed many pulsars with high dispersion measures (DMs) in the Galactic plane and some close to the GC. 

As more and more pulsars were found, the distribution of pulsars in the Galaxy could be studied in more detail. \citet{1987ApJ...319..162N} proposed that the pulsar radial density profile around the GC is Gaussian. After the completion of the first large-scale high-frequency 
pulsar survey concentrating on the Galactic plane \citep{1992MNRAS.254..177C,1992MNRAS.255..401J}, \citet{1994MNRAS.268..595J} concluded that such a model is incompatible with the observed distribution. Instead, they proposed a model in which the pulsar surface density peaks 4\,kpc from the GC. \citet{2004A&A...422..545Y} (hereafter YK04)
proposed a four-parameter pulsar radial distribution model based on the electron density model, NE2001 \citep{2002astro.ph..7156C}. \citet{2006MNRAS.372..777L} (hereafter LF06) found that the pulsar radial density profile is best described by a Gamma function. These studies have generally shown a deficit of pulsars in the inner Galaxy.

Over the last 20 years, many Galactic plane surveys have been carried out at frequencies higher than 1.4 GHz to search pulsars in the inner Galaxy especially around the GC. \citet{2006MNRAS.373L...6J} discovered two pulsars less than $\sim$40 pc from the GC with the 3.1-GHz Parkes Galactic Centre survey. 
\citet{2009ApJ...702L.177D} reported three normal radio pulsars close to Sgr A* with 2-GHz observations from the Green Bank Telescope. 
Pulsars close to the Galactic centre have also been found using space-based telescopes. NuSTAR and Swift discovered SGR~J1745$–$29 \citep{2013ApJ...770L..23M,2013ApJ...770L..24K}, which has a projected distance of only 0.097\,pc from Sgr\,A* \citep{2015ApJ...798..120B}. This source is also a radio-emitting magnetar \citep{2013Natur.501..391E} and likely to be associated with the GC region. The discovery of these pulsars enables us to investigate the pulsar distribution in more detail, especially in the GC region. 

Many factors affect the sensitivity of pulsar searches. These effects are often classified into two simplified categories. One is related to distance and the ISM, known as the distance selection effect, includes scattering, dispersion, and pulsar flux densities. The other is related to Galactic latitude and longitude, and is known as the directional selection effect, in which the sky background temperature is the most significant.

In the last decade, there has been great progress in the study of the Galactic background radiation and the Galactic electron-density model, which allows us to better understand the spatial distribution of pulsars in the Galaxy. In this paper, the pulsar distance $r$ is calculated from the Galactic electron density model, YMW16 \citep{2017ApJ...835...29Y}. A new high-resolution sky temperature map \citep{2015MNRAS.451.4311R} is used in the modeling of the directional selection effect. Contamination from extragalactic radio sources and reduced large-scale striations have been removed from this sky temperature map. We used the latest available version of the ATNF Pulsar Catalogue \citep{2005AJ....129.1993M} to obtain data on pulsar flux densities at 1.4~GHz. An empirical method based on that described by YK04 is applied to estimate selection effects. 

We describe the pulsar sample and and the process of data pre-processing in Section 2. In Section 3, the flux density distribution of pulsars at 1.4\,GHz is modelled and a new directional selection model is established. We derive the radial distribution of pulsars in Section 4 and discuss our results and conclusions in Section 5.

\section{Pulsar surface density in the Galaxy}\label{sec:surfaceDensity}

As mentioned above, the number of pulsars in the ATNF Pulsar Catalogue currently exceeds 3,300. Here, we are interested in the distribution of normal radio pulsars. Therefore, we excluded pulsars that are known to be in globular clusters, binary systems, recycled (P$<$0.03 s), and extragalactic pulsars.
In order to obtain an accurate spatial distribution of pulsars, it is better to select pulsars discovered by surveys with similar sensitivity. For this reason, pulsars discovered by four pulsar surveys with the Parkes ``Murriyang" radio telescope including the Parkes \citep{2001MNRAS.328...17M}, Swinburne \citep{2001MNRAS.326..358E}, high-latitude \citep{2006MNRAS.368..283B}, and the Perseus Arm \citep{2013MNRAS.429..579B} multibeam pulsar surveys (collectively, these four surveys are called MBPS), are selected for our working sample. A total of 1301 pulsars are located in the region of Galactic longitude $-160^{\circ} \le l \le 50^{\circ}$. The closest pulsar to the GC that was discovered by the PMPS was PSR~J0932$-$5329, at a distance 4.23 kpc. These 1301 pulsars are used to calculate the Galactic pulsar surface density.
In addition, six pulsars around the GC discovered by other pulsar surveys are included when calculating the surface density of pulsars in the GC region. See Section 4 for details.

\begin{figure*}
\includegraphics[width=18cm]{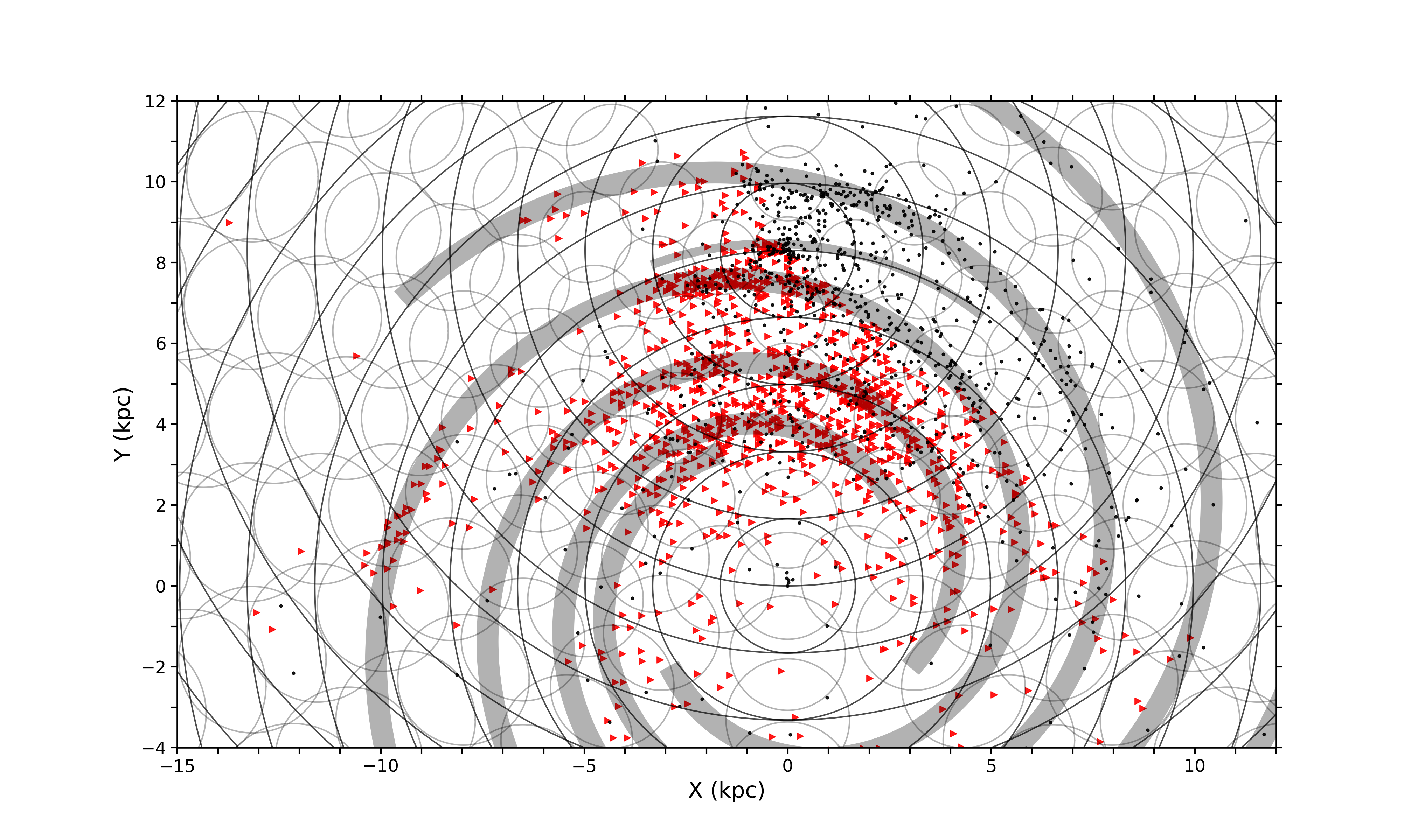}
\caption{Projection of the sample pulsars on to the Galactic plane. Pulsars discovered by the MBPS and other surveys are marked as red triangle signs and black dots, respectively. The GC is the origin of the coordinate system and the coordinates of the Sun are (0.0, 8.3). Other lines and symbols are described in the text. The spiral arms are based on parameters from \citet{2014A&A...569A.125H}.}
\label{Pulsargrid}
\end{figure*}

Pulsars discovered by the MBPS and the other surveys are projected on to the Galactic plane using different markers, as shown in Figure \ref{Pulsargrid}. Pulsar distances were estimated using the dispersion measure and the YMW16 model. The distance between the GC and the Sun was taken as $R_{\odot} = 8.3 \pm 0.23$ kpc \citep{2011AN....332..461B}. When calculating the pulsar surface density, it is necessary to grid the Galactic plane. The gridding method is as follows:

\begin{enumerate}
  \item  Concentric circles are drawn around the GC and around the Sun. The interval between adjacent circles is set at $\Delta R = R_{\odot}/5 = 1.66$ kpc. 
  \item Grid circles are centered at the intersections of these two sets of concentric circles.
  \item Distant pulsars are more difficult to detect, and the observed surface density of pulsars decreases rapidly with increasing distance. At large distances, it is inevitable that some grid circles will be empty. Increasing the grid size can help to reduce this effect. The radius of the grid circles increases with the increasing distance from the Sun according to:
  \begin{equation}
      R_{gi} = \xi(i-1)+\eta\Delta R
  \end{equation}
 where $i$ is the circle number from the Sun and ranges from 1 to 12, $\xi = 0.098$, $\eta=0.83$.
\end{enumerate}

A larger grid size will reduce the number of points with zero surface density, but it will also reduce the total number of grid points. On the other hand, a larger number of grid points will lead to a more accurate depiction of the pulsar distribution, but it will also lead to more grid points with zero surface density. To balance the number of grids and minimize the loss of pulsar surface density, we conducted multiple sets of tests to determine the optimal parameters $\Delta R$, $\xi$, and $\eta$ for the grid partitioning method described above. We selected the optimal parameters to minimize the loss of pulsar surface density while maintaining a sufficient number of grids.
After obtaining the initial surface density, which is based on these optimal parameters, there are still grid points with zero surface density, even when employing larger grid sizes for distant regions. Furthermore, there are fluctuations in pulsar surface density near spiral arms, which are likely attributable to the influence of the Galactic structure. To reduce the impact of these factors, it is necessary to apply smoothing to the initial surface density. Therefore, we smooth the pulsar surface density to reduce the fluctuations, using
\begin{equation}
{\rho_{ij}' = (\rho_{1}+\rho_{2}+4\rho_{ij}+\rho_{3}+\rho_{4})/8},
\label{eq:grid1}
\end{equation}
where $\rho_{ij}'$ is the smoothed surface density at the intersections of i$_{th}$ and j$_{th}$ circles from the Sun and the GC, and $\rho_{1},\ \rho_{2},\ \rho_{3}$, and $\rho_{4}$ are the surface density in four new grid circles at coordinates ($x_{ij}$, $y_{ij} $ + 1.66), ($x_{ij}$, $y_{ij} $ - 1.66), ($x_{ij}$ - 1.66, $y_{ij} $) and ($x_{ij}$ + 1.66, $y_{ij}$), respectively. We also averaged over the grid circles, normally six in number, surrounding a given grid point: 
\begin{equation}
\resizebox{.8\hsize}{!}{$\rho_{ij}' = (\rho_{a1}+\rho_{a2}+6\rho_{ij}+\rho_{a3}+\rho_{a4}+\rho_{a5}+\rho_{a6})/12$},
\label{eq:grid2}
\end{equation}
where $\rho_{a1}$, $\rho_{a2}$, $\rho_{a3}$, $\rho_{a4}$, $\rho_{a5}$, and $\rho_{a6}$ are the observed surface density in the six grid circles adjacent to the ($i$,$j$) grid circle. The surface density for the local 
($r \le 1.66$\,kpc) grid circle is not smoothed. 

Table \ref{tab:obsDensity} gives the average observed surface densities (in units of kpc$^{-2}$) of pulsars in each grid circle as functions of the galactocentric radius ($R$) and the distance from the Sun ($r$). 

\begin{table*}
\centering
\caption{The average observed surface density of pulsars across the Galactic plane (kpc$^{-2}$). R is the distance from the Galactic Center and r is the distance from the Sun.}
\label{tab:obsDensity}
\begin{tabular}{c|cccccccccccc}
\hline
\hline
\diagbox[height=1cm,width=3.8cm]{$r$ (kpc)}{$R$ (kpc)}&0&1.66&3.32&4.98&6.64&8.3&9.96&11.62&13.28&14.94&16.6&18.26\\ %
\hline

0&&&&&&43.89\\

1.66&&&&&40.05&17.3&13.63\\

3.32&&&&39.48&10.98&11.60&6.48&0.45\\

4.98&&&31.07&17.47&7.77&3.45&2.22&1.74&0\\

6.64&&13.58&8.79&9.62&6.74&1.79&0.38&0.87&0.23&0\\

8.3&3.04&3.19&4.68&4.47&3.2&2.09&0.44&0.13&0.11&0.02&0\\

9.96&&1.68&2.79&2.03&1.95&1.99&1.38&0.21&0.02&0&0&0\\

11.62&&&1.56&1.74&1.00&1.11&2.34&0.43&0.09&0.01&0.01&0\\

13.28&&&&1.65&0.99&0.74&1.10&0.60&0.05&0.02&0.07&0.01\\

14.94&&&&&1.13&0.65&0.31&0.27&0.13&0.01&0.01&0.01\\

16.6&&&&&&0.53&0.24&0.12&0.14&0.03&0&0\\

18.26&&&&&&&0.28&0.09&0.03&0.04&0.06&0.01\\
\hline
\hline
\end{tabular}
\end{table*}

\section{Selection effects and correction method}\label{sec:selEffects}

In studies of pulsar populations, selection effects are a major source of error, as the observed sample represents a biased subset of the entire Galaxy's pulsar population. To uncover the true nature of pulsar populations, it is crucial to identify and account for selection effects and establish a robust model.

Numerous factors contribute to selection effects, primarily including characteristics of the pulsar population, the distance to observed sources, telescope sensitivity, observation frequency, Galactic background radiation, and dispersion measure. In addition to the effects of observing frequency and instrumental characteristics, the Galactic background radiation and dispersion measure significantly influence pulsar searches \citep{2004A&AT...23..503A}.

The Galactic background radiation exhibits significant variations along different lines of sight, and its influence can be described as a directional selection effect.
In specific directions, the dispersion measure varies as a function of distance, thereby allowing the selection effect induced by dispersion measure to be approximately superimposed on the direction-related impact.

In this context, we define the fraction of missed pulsars due to a specific selection effect as $(1-K)$, where $K$ represents the selection bias factor. Consequently, the observed pulsar density, denoted as $\rho_o$, is a function of $K$ and the actual pulsar density, $\rho$, expressed as follows: $\rho_o = K \times \rho$. Assuming that the Galactic background radiation, denoted as $K(l)$, and the distance-dependent selection effect, denoted as $K(r)$, are uncorrelated, the observed pulsar surface density and the selection effect can be defined as follows:
\begin{equation}
\rho_o(r,R) = K(l)K(r)\rho(r,R),
\label{eq:sel}
\end{equation}
where $l$ is the Galactic longitude associated with the galactocentric distance ($R$) and heliocentric distance ($r$),
and $\rho_o(r,R)$ and $\rho(r,R)$ are the observed and true distributions of pulsar surface densities on the Galactic plane. The Galactic longitude is a function of $r$ and $R$ (and Galactic quadrant) and, since we are considering only densities projected on to the Galactic plane, we can take the Galactic latitude $b$ to be zero. Table \ref{tab:obsDensity} is a numeric form of $\rho_o(r,R)$.

The survey sensitivity can be described using the relation given by \citet{2020MNRAS.497.4654M}:
\begin{equation}
S_{\rm min}= \frac {\alpha \beta (T_{\rm sys} +T_{\rm sky})}{\mc{E} G \sqrt{N_{\rm p}\,\Delta\nu\,t_{\rm obs}} }\left (\frac{W_{\rm  eff}}{P-W_{\rm  eff}}\right )^{\frac{1}{2}},
\label{eq:smin}
\end{equation}
where $S_{\rm  min}$ is minimum detectable flux density for a radio pulsar; $\alpha$ is the required signal-to-noise ratio for a detectable pulsar; $\beta$ is a degradation factor relating to both non-ideal instrumental responses and offset from the beam center; $T_{\rm  sky}=T_{\rm  cmb}+T_{\rm  gal}$; $T_{\rm  sys}$ is the system temperature (K); $T_{\rm  gal}$ is the background sky temperature; $T_{\rm  cmb}$ is the cosmic microwave background (CMB) temperature; $\mc{E}$ is the search efficiency factor; $G$ is the gain of the telescope (K/Jy); $N_{\rm p}=2$ is the number of polarizations; $t_{\rm  obs}$ is the observation time per pointing (s); $\Delta\nu$ is bandwidth (Hz); $P$ is the period of a pulsar (s) and $W_{\rm  eff}$ is the effective pulse width (s). $S_{\rm min}$ determines the number and distribution of pulsars detected by any given survey. By effectively inverting its effects, the underlying or ``true'' distribution of pulsars in the Galaxy can be determined.

The longitude-dependent selection factors can be obtained from:
\begin{equation}
K(l) = 1-\int_{0}^{S_{\rm min\ (l,\ b)}}f(S_{1400})\,{\rm d}S_{1400} ,
\label{eq:selFac}
\end{equation}
where $f(S_{1400})$ is the flux density distribution function at 1400 MHz. The minimum detectable flux density can be derived from the following equation:
\begin{equation}
{S_{\rm  min \ (l,\ b)} = S_{\rm  min0}\left( 1+\frac{T_{\rm  sky,\nu}-T_{\rm  cmb}}{T_{\rm  cmb}+T_{\rm  sys}}\right)},
\label{eq:SinSelect}
\end{equation}
where $S_{\rm  min0}$ is the minimum detectable flux density for a radio pulsar considering only the CMB and radiometer noise. The pulsar with the lowest flux density among those discovered in the MBPS is J1822-0848, with an S1400 $\sim 0.04$~mJy.
Both Parkes \citep{2001MNRAS.328...17M} and the Perseus Arm \citep{2013MNRAS.429..579B} multibeam pulsar surveys cover regions with $b \le 5^{\circ}$ with similar observation time per point. Considering the variation of background radiation with Galactic latitude, the majority of survey areas exhibit similar sensitivity.
Therefore, we set $S_{\rm  min0}$ to be 0.04 mJy; $T_{\rm  sys}$ is the telescope system temperature, $T_{\rm  sys}$ = 21 K for the Parkes multibeam receiver \citep{2001MNRAS.328...17M}. Equation (\ref{eq:smin}) indicates the background radiation of the Galaxy will lead to increase of the minimum detectable flux density for a pulsar. Therefore, we divide the density of pulsars in each grid circle by its corresponding K($l$).

$K(l)$ mainly depends on the effective temperature of the sky background radiation, $T_{\rm sky}$. 
The sky temperature at 408 MHz can be obtained from \citet{2015MNRAS.451.4311R} and scaled to the observing frequency (1374\;MHz) through the following relation:
\begin{equation}
{T_{{\rm  sky},\nu} = (T_{{\rm  sky},\nu_0} - T_{\rm  cmb})\,\left(\frac{\nu}{\nu_0}\right)^{-2.6}+T_{\rm  cmb}},
\label{eq:tsky}
\end{equation}
where $T_{{\rm sky},\nu_0}$ is the sky temperature at 408 MHz; $T_{\rm  cmb}$ = 2.275\;K \citep{2009ApJ...707..916F}.

\begin{figure*}
\includegraphics[width=18cm]{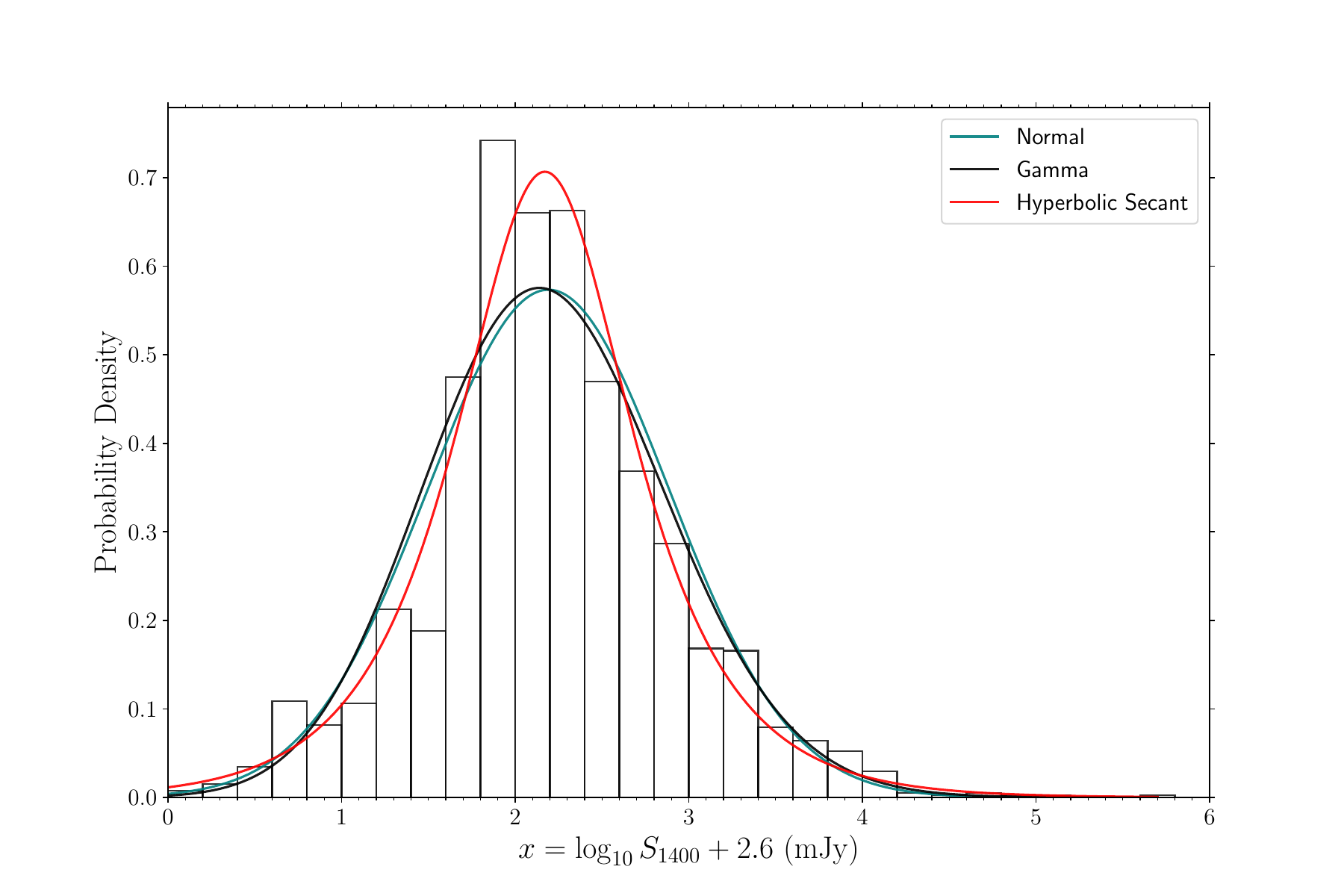}
\caption{Flux density distribution of normal pulsar at 1400 MHz. The best-fit hyperbolic secant is shown in the red curve. In addition we fit the distribution with normal (green) and gamma (black) distributions, which do not describe the observed distribution well.}
\label{fig:S1400}
\end{figure*}

\begin{figure*}
\includegraphics[width=18cm]{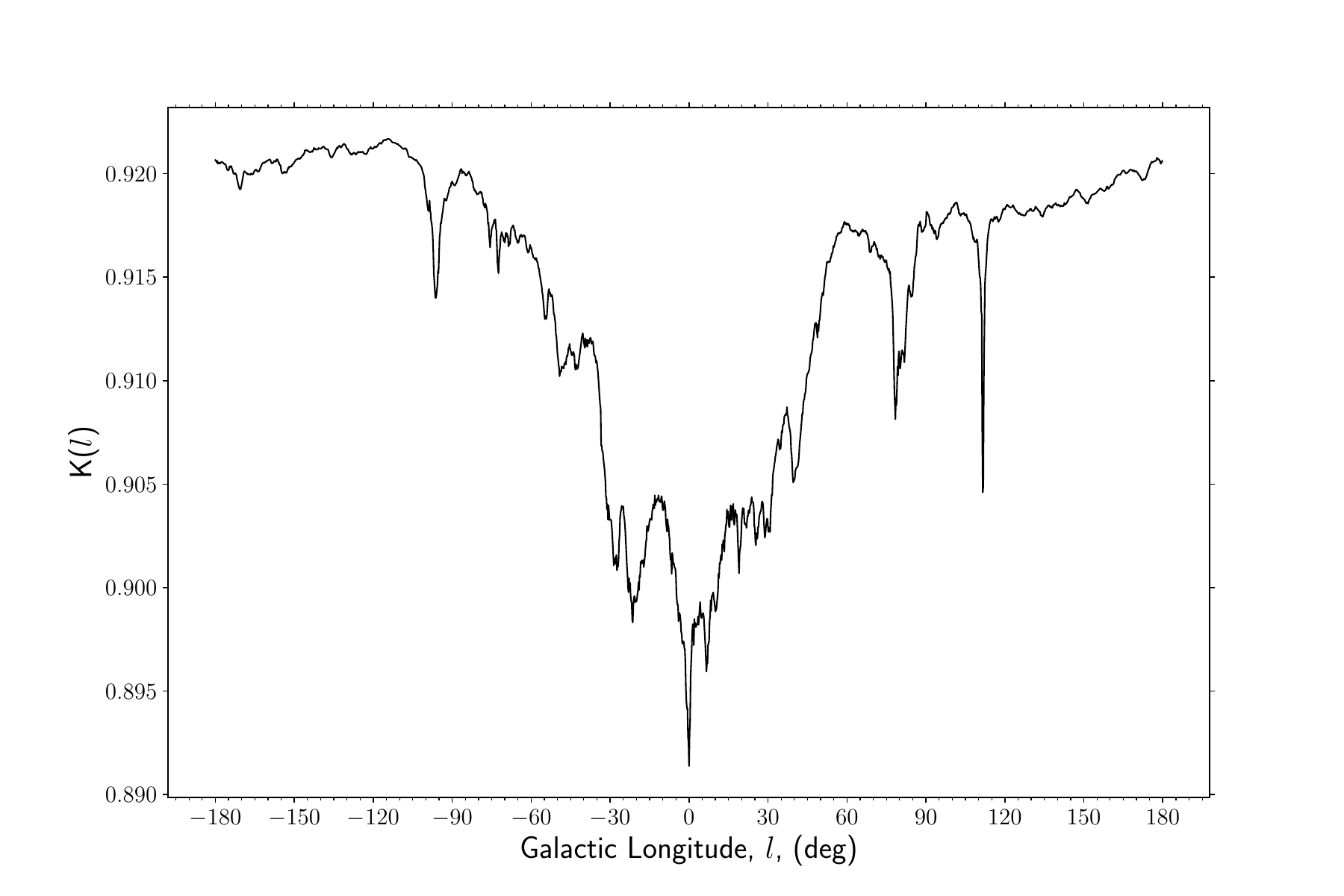}
\caption{Directional selection correction factor K($l$) caused by Galactic background radiation. }
\label{fig:gl_sel}
\end{figure*}

Figure~\ref{fig:gl_sel} shows the longitude dependence of $K(l)$ obtained by integrating $T_{\rm sky}$ over $\pm 5^\circ$ of Galactic latitude $b$ and substituting into Equation~\ref{eq:smin}. 
In the calculation of K($l$), YK04 assumed the flux densities of normal pulsars to be uniformly distributed, which may lead to an underestimation of the proportion of low-flux pulsars.

We model the pulsar flux density distribution using 2616 normal pulsars. The pulsar flux densities at 1400\,MHz are fitted using logarithmic-normal, gamma and hyperbolic-secant distributions, as shown in Figure \ref{fig:S1400}.
The pulsar flux density distribution clearly is better fitted by the hyperbolic-secant function. The flux density distribution of normal pulsars at 1400\,MHz is expressed as:
\begin{equation}
{f(x; \mu, \sigma)=\frac{1}{\pi\sigma}{\rm sech}(\frac{x-\mu}{\sigma})},
\label{eq:fluxdist}
\end{equation}
where $\mu = 2.17\pm0.03$ , $\sigma = 0.45\pm0.02$, $x = \log_{10} (S_{\rm  1400})+2.6$. Here, 2.6 is specifically chosen to ensure that $x$ remains positive for the observed sample. 
Based on the results of Kolmogorov-Smirnov test, the hyperbolic-secant function is the best fit to the observed pulsar flux density distribution at 1400MHz. Then, the pulsar flux distribution is used to estimate the effects of radiometer noise and Galactic background radiation. According to Equation (\ref{eq:smin}) to Equation (\ref{eq:fluxdist}), the longitude-dependant selection factor can be written as:
\begin{equation}
{K(l) =1- \int_{0}^{\log_{10} S_{\rm  min \ (l,\ b)} +2.6} f(x; \beta, \alpha) {\rm d}x}.
\label{eq:kl_cal}
\end{equation}
The directional selection-correcting factor K($l$) is calculated from Equation (\ref{eq:smin}) $\sim$ Equation (\ref{eq:kl_cal}), as shown in Figure \ref{fig:gl_sel}.

\citet{1974PASJ...26..255K} proposed a method to calculate the distance selection effect for supernova remnants (SNRs). It was later used to study the distribution of pulsars in the Galaxy \citep[see e.g.][]{1980AcASn..21..237H,1981ATsir1164....1Y,1984AZh....61..708G,1989AcApS...9..232W,2004A&A...422..545Y}. We adopt this method to calculate the distance-dependent selection factor. Whether a flux-density-limited survey can detect a pulsar depends on its luminosity. For a given distance ($d$), the minimum detectable luminosity $L_{\rm  min}$ can be scaled to $S_{\rm  min}$. The relationship is as follows:
\begin{equation}
{L_{\rm  min} = S_{\rm  min}d^2},
\label{eq:luminity}
\end{equation}
where $S_{\rm  min}$ can be derived from Equation (\ref{eq:smin}).

$S_{\rm  min}$ is also related to the pulse width. 
Pulse broadening, scattering and scintillation will reduce the sensitivity of pulsar surveys. These factors strongly depend on distance and reduce the detected number of pulsars.
We simply assume that the combined effect of these factors can be described by the exponential law as:
\begin{equation}
\resizebox{.8\hsize}{!}{$K(r)=K_{1}(r)K_{2}(r)\cdot \cdot \cdot K_{i}(r)=e^{-c_{1}r}e^{-c_{2}r}\cdot \cdot \cdot e^{-c_{i}r}=e^{-cr}$},
\label{eq:selectkr}
\end{equation}
where $K_{1}(r)$, $K_{2}(r)$, etc. are distance-dependent correction factors resulting from scintillation, scattering, and other effects. The functional form of the distance selection effect conveniently represents the exponential decay of a transmitted electromagnetic wave. It is difficult to quantitatively estimate each of these factors separately. We use the data in Table \ref{tab:obsDensity} to estimate their combined effect as described below.

\begin{figure*}
\includegraphics[width=18cm]{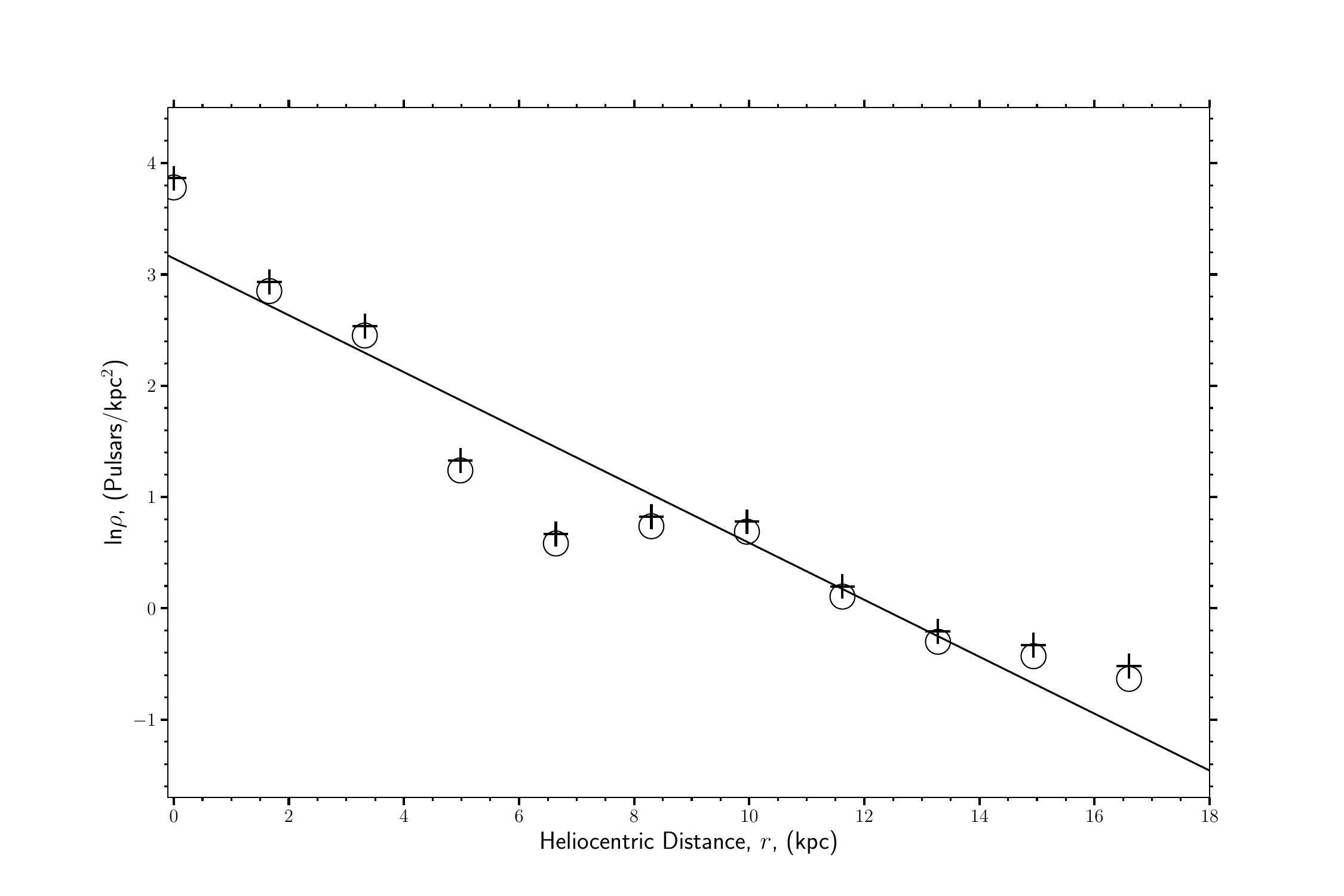}
\caption{ Pulsar surface densities on the galactocentric circle $R_{\odot}$ = 8.3 kpc as a function of distances from the Sun (Col. 6 in Table \ref{tab:corrdensity}). Apparent densities are marked by circles and corrected densities due to the direction-dependent selection effect are marked by pluses. The solid line is an LMS fitting of them.}
\label{fig:kr}
\end{figure*}

We assume that the surface density of pulsars is symmetrically distributed around the GC. Then the surface density at a distance of $R_{\odot}$ from GC can be calculated using Equation (\ref{eq:sel}),
where $\rho(r,R_{\odot})$, the apparent density at R = 8.3 kpc, is shown in Col.~6 of Table \ref{tab:obsDensity}. 

 Figure \ref{fig:kr} shows the observed and corrected surface density of pulsars as a function of the heliocentric distance $r$. 
 $K(r)$ and $\rho_{o}(r,R_{\odot})$ can be derived by fitting the data in Figure \ref{fig:kr} with a simple exponential:
\begin{equation}
{K(r)=\exp(-cr)}.
\label{eq:krf}
\end{equation}
 
 $\rho_{o}(r,R_{\odot})$ is constant and K($l$) is known. The local surface density of pulsars $\rho_{o}(r,R_{\odot})$ within $r \le 0.83$~kpc can be estimated using all the available data, taking into account the different survey sensitivities as a function of Galactic longitude and latitude. 
 We obtained 46.8 pulsars kpc$^{-2}$ with 1400-MHz flux densities greater than $S_{min,1400}$ as marked by plus signs in Figure \ref{fig:kr}. 
K(r) and $\rho_{o}(r,R_{\odot})$ are derived by a least-mean-squares (LMS) fitting of the corrected surface densities shown in Figure \ref{fig:kr} with:
\begin{equation}
\resizebox{.85\hsize}{!}{$\ln (\rho_{o}(r,R_{\odot})/K(l))= \ln(\rho(0,R_{\odot})K(r))=\ln \rho(0,R_{\odot}) -cr$},
\label{eq:lnkr}
\end{equation}
giving $\ln \rho(0,R_{\odot})= 3.16 \pm 0.26$ and $c = 0.248 \pm 0.027$. 

\begin{figure*}
\centering
\includegraphics[width=18cm]{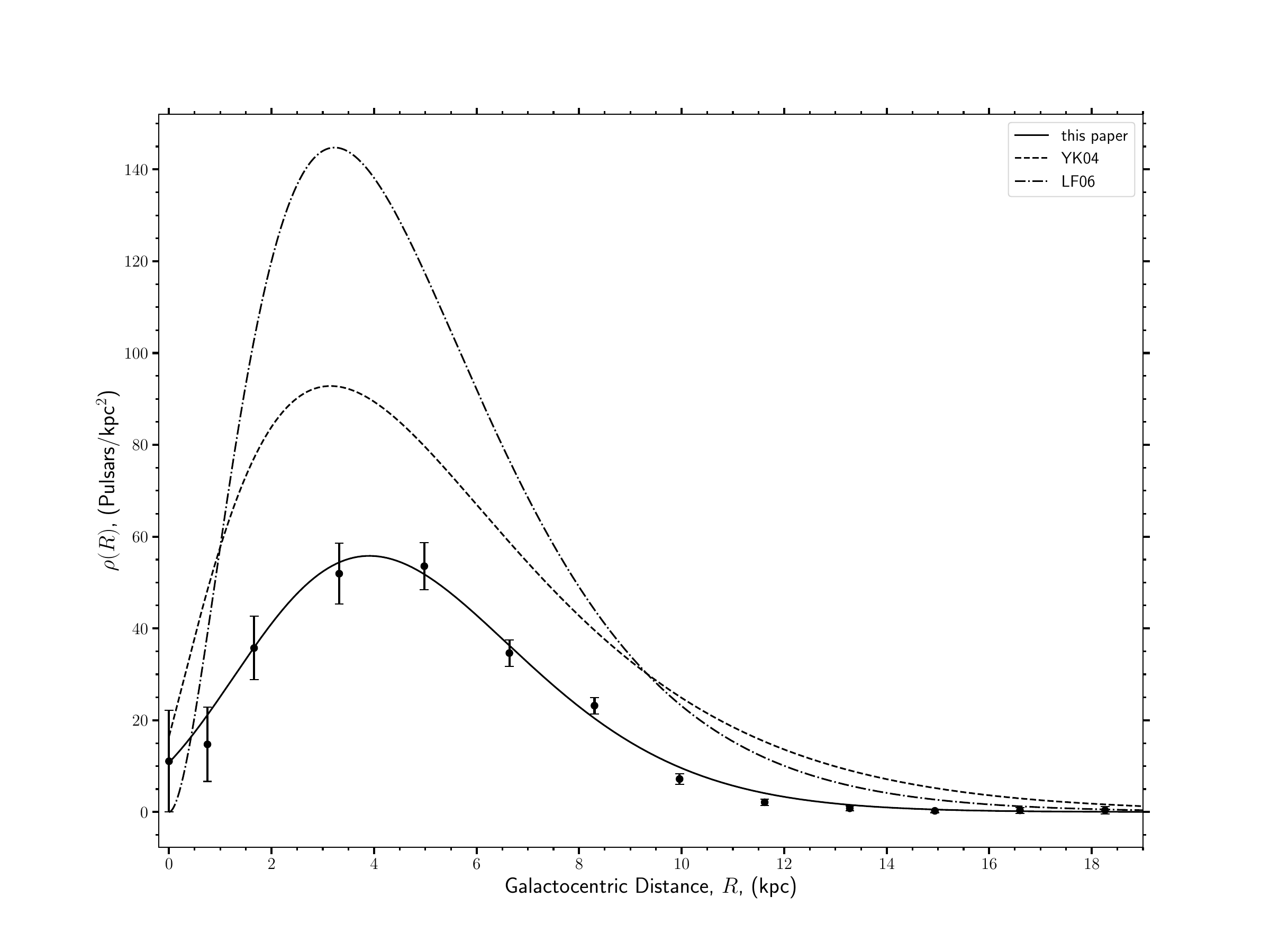}
\caption{ Radial distribution of pulsar surface density. The error bars are derived from Equation (\ref{eq:densityerror}). Dashed and dotted lines represent the radial distribution functions fitted by YK04 and LF06, respectively.}
\centering
\label{fig:PsrRdist}
\end{figure*}

The corrected surface density of pulsars is given in Table \ref{tab:corrdensity} using $K(l)$ and $K(r)$ derived from Equation (\ref{eq:kl_cal}) and Equation (\ref{eq:krf}). The local surface density of pulsars is $\rho(0,R_{\odot}) = 24^{+7}_{-5}$ pulsars kpc$^{-2}$ obtained by fitting the surface density shown in Figure \ref{fig:PsrRdist}. The true local pulsar density is higher than the fitted value, which may be accounted for in several different ways. One is a possible over-estimation of the electron density near the Sun by the YMW16 model. There appears to be a similar problem with the Carina arm, which would affect the apparent and corrected densities on or near the Solar Circle at distances of 5-8 kpc, and hence also the derived radial distribution.

\begin{table*}
\centering
\caption{Corrected surface densities of pulsars.}
\label{tab:corrdensity}
\begin{tabular}{c|cccccccccccc}
\hline
\hline
\diagbox[height=1cm,width=3.8cm]{$r$ (kpc)}{$R$ (kpc)}&0&1.66&3.32&4.98&6.64&8.3&9.96&11.62&13.28&14.94&16.6&18.26\\ %
\hline
0.0&&&&&&48.63&&&&&&\\
1.66&&&&&84.62&28.42&22.35&&&&&\\
3.32&&&&125.9&28.01&28.93&16.02&1.11&&&&\\
4.98&&&149.55&68.09&29.42&13.27&8.33&6.5&0.0&&&\\
6.64&&98.66&53.52&55.49&38.9&10.18&2.15&4.91&1.3&0.0&&\\
8.3&33.33&29.06&43.12&39.02&28.19&17.98&3.79&1.11&0.94&0.17&0.0&\\
9.96&&27.8&37.69&27.35&25.68&25.91&17.88&2.72&0.26&0.0&0.0&0.0\\
11.62&&&38.97&36.54&19.97&21.98&45.94&8.5&1.75&0.2&0.19&0.0\\
13.28&&&&62.21&30.64&22.15&33.56&17.75&1.49&0.59&2.06&0.29\\
14.94&&&&&64.31&30.09&14.01&12.22&5.83&0.45&0.44&0.44\\
16.6&&&&&&45.52&17.13&8.2&9.48&2.02&0.0&0.0\\
18.26&&&&&&&36.3&9.62&3.1&4.1&6.09&1.02\\
\hline
\hline
\end{tabular}
\end{table*}


The surface density of pulsars as a function of radial distance can be derived by averaging the columns in Table \ref{tab:corrdensity}.
The uncertainty of the pulsar surface density can be calculated as follows \citep{2004A&A...422..545Y}:
\begin{equation}
{\frac{\sigma_{\rho_{ijc}}^2}{\rho_{ijc}^2}=\left(\frac{\sqrt{N_{ij}}}{N_{ij}}\right)^2+(r\sigma_{c})^2},
\label{eq:error}
\end{equation}
where $\sigma_{\rho_{ijc}}$ is the error of $\rho_{ijc}$, and $\sigma_{c}=0.027$ is the standard deviation calculated using Equation (\ref{eq:lnkr}). The surface density uncertainty in each grid is shown in Table \ref{tab:densityerror}. 
We can obtain the radial distribution of pulsars in the Galaxy by averaging the corrected pulsar surface density in Table \ref{tab:corrdensity}.
As the inferred surface density of pulsars within individual grid circles tends towards zero, the formal uncertainty also tends towards zero. Physically, the uncertainty within these grid circles can be more accurately determined by averaging the uncertainties of neighboring circles. However, when computing the radial distribution function, grid circles with a surface density of zero also possess a weight of zero. Therefore, we did not calculate their uncertainty by averaging the uncertainties of adjacent grid circles.

\begin{table*}[t!]
\centering
\caption{Errors of surface densities of pulsars}
\label{tab:densityerror}
\begin{tabular}{c|cccccccccccc}
\hline
\hline
\diagbox[height=1cm,width=3.8cm]{$r$ (kpc)}{$R$ (kpc)}&0&1.66&3.32&4.98&6.64&8.3&9.96&11.62&13.28&14.94&16.6&18.26\\ %
\hline
0.0&&&&&&4.8&&&&&&\\
1.66&&&&&8.97&4.34&3.81&&&&&\\
3.32&&&&15.77&5.28&5.34&3.75&0.92&&&&\\
4.98&&&24.2&12.27&6.61&4.0&3.02&2.62&0&&&\\
6.64&&21.58&12.71&12.93&9.82&3.96&1.66&2.58&1.27&0&&\\
8.3&11.07&9.53&12.88&11.77&9.23&6.67&2.58&1.34&1.23&0.51&0&\\
9.96&&11.34&13.54&10.6&10.06&10.1&7.73&2.47&0.73&0&0&0\\
11.62&&&16.86&15.41&9.72&10.38&18.24&5.51&2.24&0.75&0.71&0\\
13.28&&&&28.0&15.38&12.0&16.43&10.23&2.39&1.47&2.82&1.02\\
14.94&&&&&32.71&17.28&10.03&9.18&5.82&1.49&1.46&1.46\\
16.6&&&&&&28.21&13.33&8.24&8.96&3.74&0&0\\
18.26&&&&&&&27.04&10.6&5.51&6.39&7.94&3.06\\
\hline
\hline
\end{tabular}
\end{table*}

In Figure \ref{fig:PsrRdist}, the solid line corresponds to the fitted radial distribution (Equation \ref{eq:densityfunc}).
The surface densities for each column in Table~\ref{tab:corrdensity} are averaged with the weight of each grid based on its uncertainty and calculated as follows:
\begin{equation}
\resizebox{.8\hsize}{!}{$\rho(R_{j})=\frac{\sum _{i}\frac{\rho_{ijc}}{\sigma_{ijc}^2}}{\sum_{i}1/\sigma_{ijc}^2}\ \ \ \ {\rm and}\ \ \ \     \sigma^2(R_{j})=\left( \sum_{i}(\sigma_{ijc}^{-2}) \right)^{-1}$},
\label{eq:densityerror}
\end{equation}
where $j$ and $i$ are the serial numbers of column and row; $R_{j}$ is distance from the GC corresponding to the first row in Tables 1 -- 3. As shown in Table \ref{tab:corrdensity}, the surface density at GC ($R < 1.2$ kpc) is about 33 pulsars $\mathrm{kpc^{-2}}$ considering only MBPS pulsars and its uncertainty is $\sim$11 pulsars $\mathrm{kpc^{-2}}$.

We found that no pulsar was detected by the MBPS surveys within 0.53 kpc of the 
GC. Therefore, we include pulsars discovered by other surveys. There were six pulsars discovered in the GC region, and four more in the region of 0.53 kpc $< R < 1.1$ kpc from the GC. After correcting for the direction and distance selection effects, the surface density of GC region is $14.74\pm8.08$ pulsars $\mathrm{kpc^{-2}}$. These results are used to fit the radial distribution and are shown in Figure \ref{fig:PsrRdist}.

The radial distribution of the pulsar surface density shown in Figure \ref{fig:PsrRdist} can be fitted by the gamma function:
\begin{equation}\label{eq:densityfunc}
\resizebox{.8\hsize}{!}{$\rho(r)=A\left(\frac{r+R_{pdf}}{R_{\odot}+R_{pdf}} \right)^a \exp\left[ -b \left( \frac{r-R_{\odot}}{R_{\odot}+R_{pdf}}\right)\right]$},
\end{equation}
where $R_{\odot}$ = 8.3 kpc is the Sun-GC distance, A = 20.41 $\pm$ 0.31 $\mathrm{kpc^{-2}}$, a = 9.03 $\pm$ 1.08, b = 13.99 $\pm$ 1.36, $R_{pdf}$ = 3.76 $\pm$ 0.42. The surface density of pulsars increases from the GC and reaches a maximum at a galactocentric radius of $\sim$3.91 kpc. According to \citet{2014A&A...569A.125H}, the average starting position of the four spiral arms, i.e., the Perseus Arm, the Carina-Sagittarius Arm, the Crux-Scutum Arm and the Norma Arm, is 3.57 kpc, which is close to the position where the pulsar surface density reaches its peak.

\section{Discussion}

\begin{figure*}
\centering
\includegraphics[width=16cm]{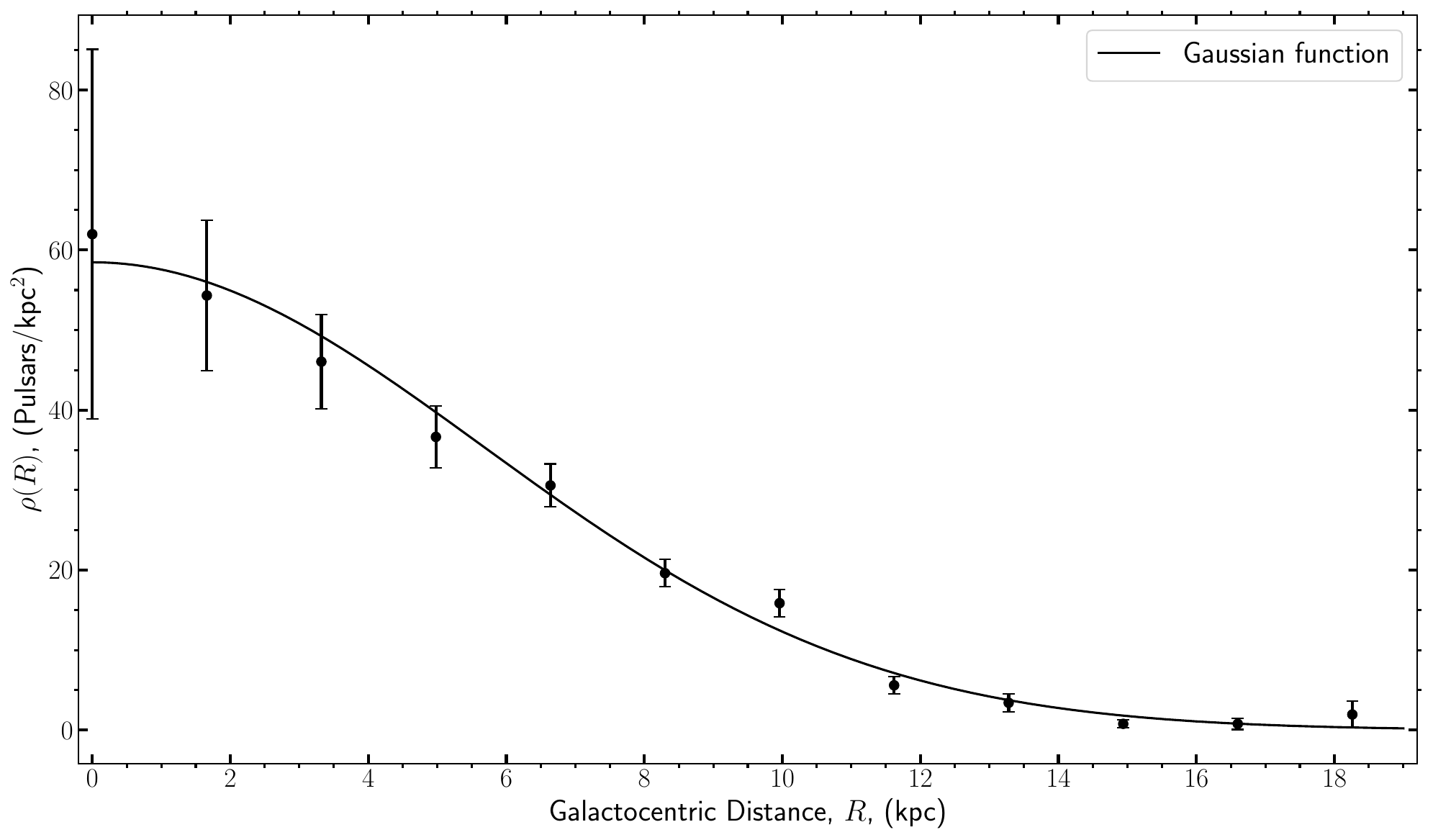}
\caption{Radial distribution of pulsar surface density from a simulated population of pulsars with Gaussion distribution proposed by \cite{1987ApJ...319..162N}. The points and error bars are derived from simulated detectable pulsars, calculated based on the methods outlined in sections 2 and 3. The solid black line represents the results of a Gaussian function fit.}
\centering
\label{fig:PsrRdistVerify}
\end{figure*}

The methods we have used to model the pulsar radial distribution can effectively reduce the selection effects when we have a large sample.
To validate the method of estimating radial distributions, we simulate a population of pulsars with the Gaussian radial distribution proposed by \cite{1987ApJ...319..162N}. Using {\sc PsrPopPy}, we generated a simulated pulsar sample from this input distribution and applied the method to the sample. It returned a distribution compatible with Gaussian distribution, as illustrated in Figure \ref{fig:PsrRdistVerify}.

A similar method was used by LF06. However, it is difficult to estimate the radial distribution in the GC region using this method since there are few known pulsars there. In contrast to the sample utilized in the studies conducted by YK04 and LF06, our sample exhibits a two-fold increase in the population of known pulsars, encompassing a substantially larger number of distant pulsars characterized by high dispersion measures. As shown in Table 1, the apparent density of pulsars has increased significantly, especially for grids at a large distance from the Sun and GC.

Distance is crucial for calculating the distribution of pulsars in the Galaxy. We make use of the YMW16 distance model. Compared to NE2001, YMW16 used more independent pulsar distances to calibrate the model, and an improved spiral-arm structure for the Galaxy \citep{2014A&A...569A.125H}. 

Sky temperature is one of the most important factors affecting survey sensitivity. In order to update the scaling relation utilized for calculating the directional selection-correcting factor, we used an improved sky temperature map that effectively mitigates contamination arising from extragalactic radio sources and minimizes the presence of large-scale striations \citep{2015MNRAS.451.4311R}. This improves on the method used in YK04, which used a continuous function to describe the sky background temperature. We also fit the flux density distribution of the pulsar to calculate directional selection-correcting factor. If, as in YK04, the pulsar flux distribution is assumed to be uniform, the number of pulsars in the GC region will be overestimated. As shown in Figure \ref{fig:S1400}, K($l$) exhibits strong fluctuations and the survey sensitivity is severely reduced around the GC. The distance-correcting factor can be calculated using K($l$).

\citet{1974PASJ...26..255K} proposed an empirical method for correcting distance selection effects to calculate the radial distribution of supernova remnants. This method was based on the assumption of central symmetry in the Galaxy.
The association of spiral arms with pulsars may render the assumption of central symmetry no longer valid. To mitigate the influence of spiral arms, we employ the smoothing method outlined in Section 2. Furthermore, the smoothing procedure enhances our adherence to the assumption of symmetry around the GC. After correcting for directional selection effects, the distance selection effect function can be obtained by fitting the pulsar surface densities at the same galactocentric distance but varying heliocentric distances. We obtain $c=0.248$ for the distance-correcting factor $K(r)=e^{-cr}$ which is smaller than that of YK04. This discrepancy primarily arises from the utilization of different electron-density models for distance estimation.

Isolated neutron stars are believed to originate from supernova events in OB-type Population I stars. Therefore, neutron stars should be located along the spiral arms. In fact, \citet{2003MNRAS.342.1299K} noted that the Galactic spiral structure is clearly visible from the distribution of young pulsars. \citet{2019MNRAS.487.1400C} also found that most OB stars are located in the Galactic plane, and clearly trace the structure of the Galactic spiral arms. 

Figure \ref{Pulsargrid} also shows that the pulsar positions are highly correlated with the spiral arm structure. However, the pulsar positions in Figure \ref{Pulsargrid} 
are entirely derived from the YMW16 distance model and are not independent evidence for real spiral concentrations. 
We use the spiral arms as the location for the birth of pulsars outside the GC, but there is still some uncertainty about it.

The results of \citet{2014A&A...569A.125H} are used to model the structure of the Galactic spiral arms. The spiral arms are described in polar coordinates with the GC at the origin. The x-, y-, and z-axes are, respectively, parallel to $(l, b) = (90^{\circ}, 0^{\circ}), (180^{\circ}, 0^{\circ})$, and $(0^{\circ}, 90^{\circ})$, in a right-handed Cartesian frame. As usual, $
x = R\cos{\theta}$ and $y = R\sin{\theta}$. There are five spiral arms, described by the following formula:
\begin{equation}
{\theta=\frac{\ln (\frac{R}{R_{0}})}{\tan(\theta_1)}+\theta_0},
\label{eq:spirtheta}
\end{equation}
where $\theta$ and $r$ are polar coordinates; $R_{0}, \theta_{0}$ and $\theta_{1}$ are the initial radii, the starting azimuth angle, and the pitch angle for the i$_{th}$ spiral arm, respectively. The parameters of each arm given in Table \ref{tab:spiralpara} are taken from \citet{2014A&A...569A.125H} and consistent with those of YMW16.

The location of a simulated pulsar on spiral arm depends on the radial distribution of pulsars and the structure of the spiral arms. The spiral structure is realized by choosing the locations of simulated pulsar so that their projections lie on arms and subsequently adjusting them to simulate a spread around the arm centroids. The number of pulsars in each spiral arm is different because of their different lengths and positions. Therefore, the probability of a simulated pulsar lie on a spiral arm is different for the different spiral arms. The number of pulsars in each spiral arm can be obtained by integrating the pulsar radial distribution function along the spiral arm. We calculate the number of pulsars in each spiral arm in the form of discrete point integral. The number of pulsars in a certain spiral arm divided by the total number of pulsars in all the spiral arm is the probability we mentioned above. The Local arm has been taken into account in the simulation. The probability can be obtained by the following equation:
\begin{equation}
\resizebox{.8\hsize}{!}{$R_{j} =\frac{\sum_{i=1}^N \sqrt{(x_{i+1}-x_{i})^2+(y_{i+1}-y_{i})^2}\rho(r_{ij})0.07r_{ij}}{\sum_{j=1}^{5}\sum_{i=1}^N \sqrt{(x_{i+1}-x_{i})^2+(y_{i+1}-y_{i})^2}\rho(r_{ij})0.07r_{ij}}$},
\label{eq:ratiospiram}
\end{equation}
where $R_{j}$ is the probability of a simulated pulsar to lie on the $j_{th}$ spiral arm, N is the number of equally spaced intervals for a spiral arm, $r_{ij}$ is the $i_{th}$ point's galactocentric radius in spiral arm $j$, $\rho(r_{ij})$ is the pulsar surface density at galactocentric radius $r_{ij}$, and $x_{i}$ and $y_{i}$ are the $i_{th}$ point’s Cartesian coordinates. 

Ratios of the number of pulsars in each spiral arm are shown in Table \ref{tab:spiralpara}. A distance $r_{\rm raw}$ from the GC is then chosen according to the pulsar radial distribution. For a pulsar to lie on the centroid of spiral arm $i$, the corresponding polar angle $\theta_{\rm raw}$ is calculated according to Equation (\ref{eq:spirtheta}). The method used in \citet{2006ApJ...643..332F} is adopted to broaden the distribution to avoid artificial features near GC. The corrected value of $\theta$ in the polar coordinate system is $\theta_{\rm corr} \exp({-0.35 r_{\rm ini}/{\rm kpc}})$, where $\theta_{\rm corr}$ is a random number uniformly distributed in the interval $[0, 2\pi]$. Finally, pulsars are spread along the spiral arm centroid. We assume $r_{\rm last}$ follows the normal distribution with mean $r_{\rm raw}$ and standard deviation $0.07r_{\rm raw}$.

\begin{figure}
\centering
\includegraphics[width=8cm]{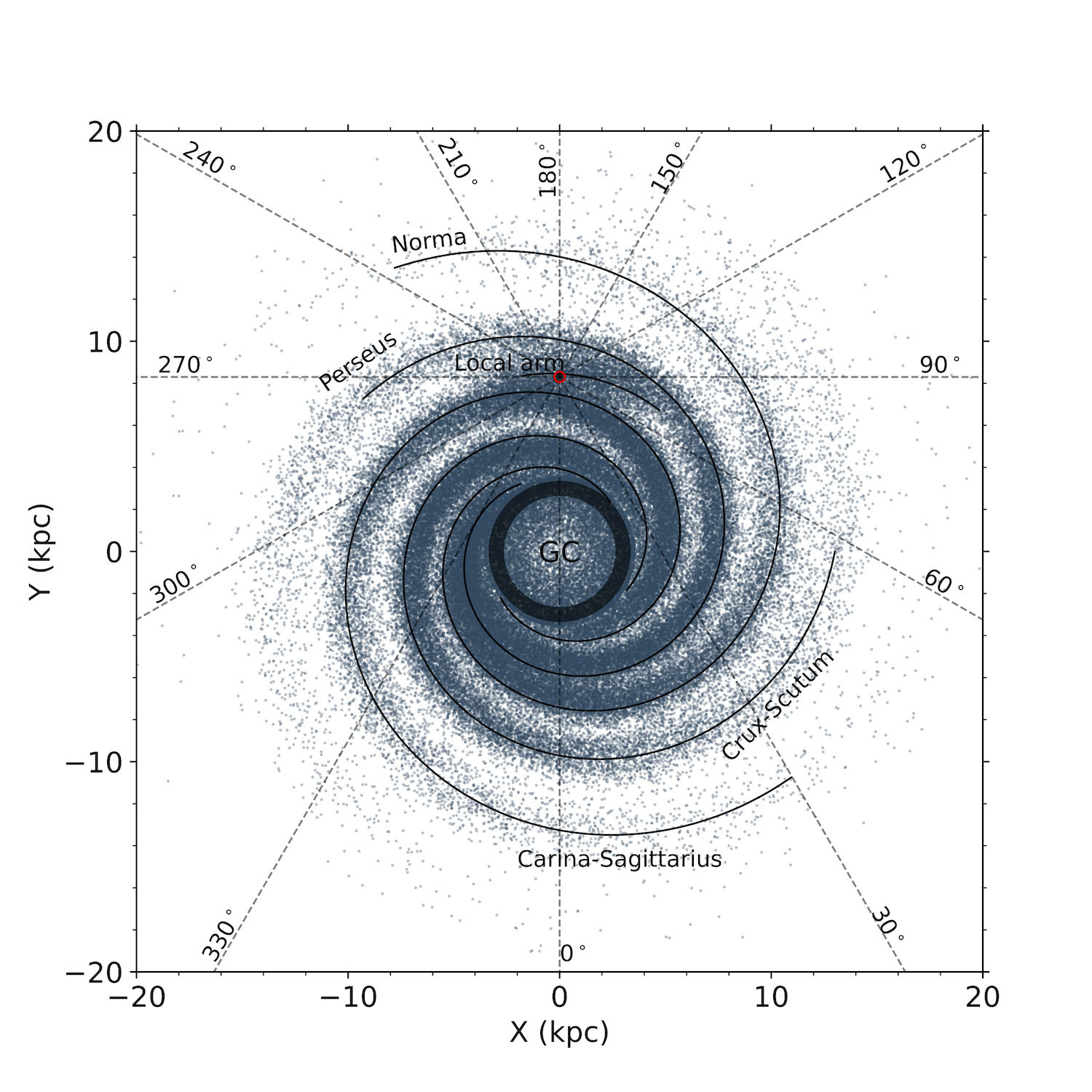}
\caption{ An example of the simulated initial distribution of pulsars. Each point represents the birth location of a pulsar projected onto the Galactic plane. The position of the Sun is indicated by the red circle. Solid lines show the spiral-arm centroids.}
\label{fig:MCPsr}
\end{figure}

For the Galactic Center region, considering the difference between it and the spiral arms, we use another method to generate the birth positions of isolated neutron stars. The distance from the Galactic Center and the Galactic longitude are generated by two independent random processes. The probability relation for the distance between the Galactic Center and the pulsar can be obtained as follows \citep{2021ApJ...916..100R}:
\begin{equation}
{P(r) \propto 2\pi R\rho(R)},
\label{eq:calpr}
\end{equation}
where $\rho(R)$ is the surface density and $R$ is the distance from the Galactic Center. The Galactic longitudes of the simulated pulsars are randomly chosen in the interval $[0, 2\pi]$~rad. Figure \ref{fig:MCPsr} shows the positions of all the simulated pulsars on the $x-y$ plane.

\begin{table*}[]
\centering
\caption{The proportion of pulsars in the composition of the Galaxy and Spiral Arm Parameters}
\label{tab:spiralpara}
\setlength{\tabcolsep}{10pt}
\begin{tabular}{cccccc}
\hline
\hline
Galactic structure Number &Name&$\theta_{0}$ (deg)&$\theta_{1}$ (deg)&$r_{0}$ (kpc)&Ratio (\%) \\
\hline
1&Norma&44.4&11.43&3.35&19.77\\
2&Perseus&120.0&9.84&3.71&20.91\\
3.&Carina-Sagittarius&218.6&10.38&3.56&21.20\\
4&Crux-Scutum&330.3&10.54&3.67&20.17\\
5&Local arm&55.1&2.77&8.21&2.87\\
6& Galactic centre &-&-&-&15.08\\
\hline
\hline
\end{tabular}
\end{table*}

\begin{figure}
\centering
\includegraphics[width=8.5cm]{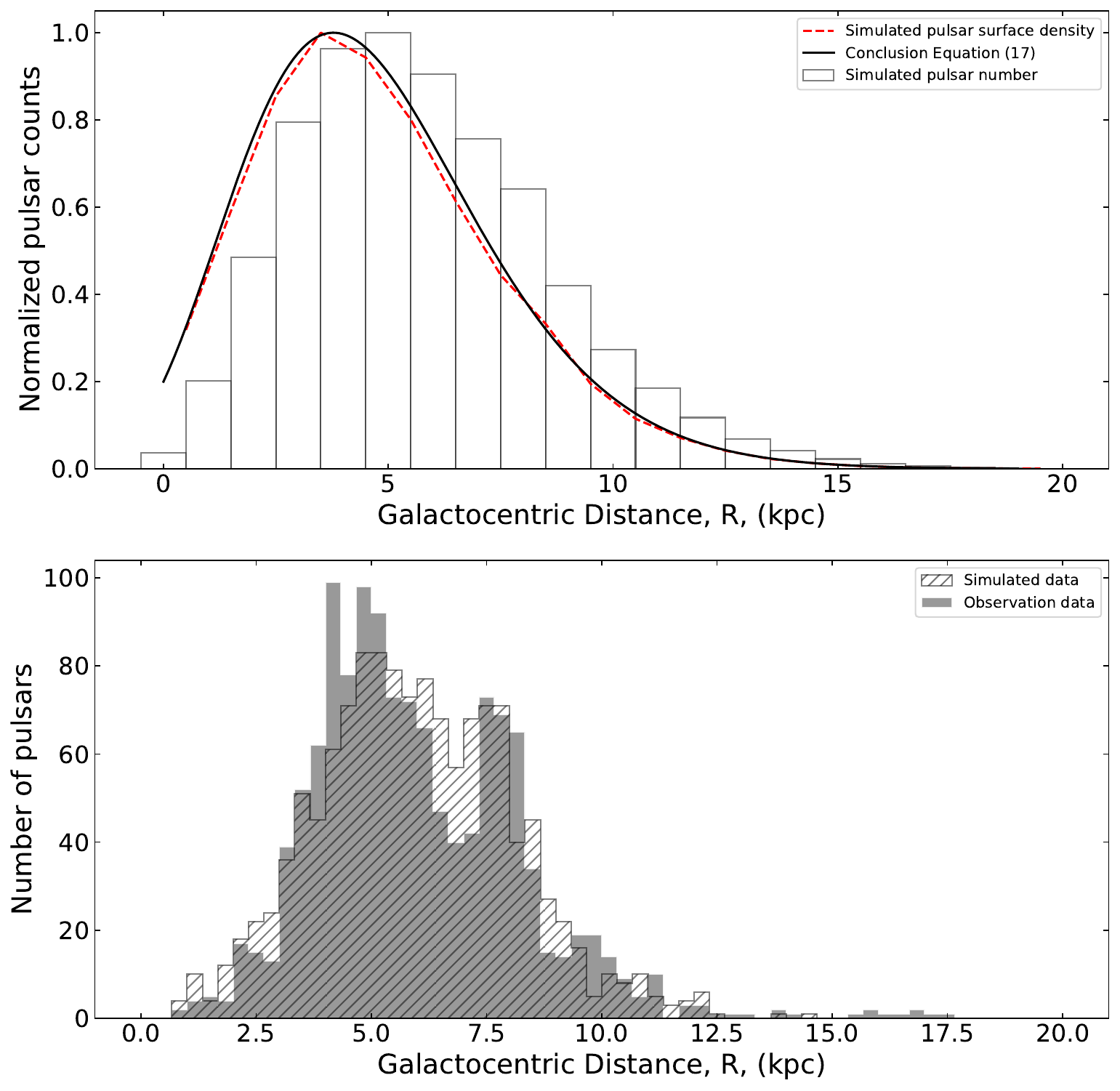}
\caption{ 
Upper: Surface density distribution of pulsar populations on the Galactic plane. The red line represents the surface density distribution of a simulated pulsar population using PsrPopPy, while the black line illustrates the formula derived from Equation \ref{eq:densityfunc}. The histogram shows the normalized distribution of simulated pulsar counts. 
Lower: histograms of the Galactic center distances for pulsars discovered in simulated MBPS survey and actual MBPS survey. The hatched histogram represents the distribution of Galactic center distances for pulsars discovered in the simulated MBPS survey, while the gray-shaded histogram represents the distribution of Galactic center distances for pulsars discovered in the actual MBPS survey.}
\centering
\label{fig:MBPSPsrRdist}
\end{figure}

We employed {\sc PsrPopPy} to simulate the MBPS survey using Equation 17 and the method for generating simulated pulsar positions described in \citet{2014MNRAS.439.2893B}. This allowed us to obtain distances from the Galactic center for pulsars discovered in the simulated MBPS survey. These simulated results were subsequently compared with the actual discoveries of the MBPS survey, as illustrated in Figure \ref{fig:MBPSPsrRdist}. It is apparent that the radial distribution of the simulated pulsar population closely aligns with the conclusions drawn from Equation \ref{eq:densityfunc}.

To assess the agreement between simulated and observed data, hypothesis tests were conducted, including the Kolmogorov-Smirnov (K-S) test and the Anderson-Darling (A-D) test.
Both of the tests indicate there is no significant difference between simulated data and observed data.

In the GC region the surface density of pulsars that we derived is higher than that of LF06 but lower than YK04, and reaches its peak at a larger distance, but the overall density is lower than that of YK04 and LF06. 
After applying the \citet{TM98} beaming correction, the pulsar surface density in the GC region that we obtained is almost the same as that obtained by \cite{2006MNRAS.372..777L} from a Monte Carlo simulation.

By integrating the radial distribution function, the estimated number of detectable pulsars in the Galaxy is $(1.1\pm 0.2) \times 10^4$, with the proportion within the Galactic center region ( $R \le 3.57$ kpc) being about 15\%. 

Only a small fraction of pulsars have their radiation beams directed towards Earth.
Beam selection effects are particularly important to consider when calculating the birth rate of pulsars in the Galaxy. 
According to \citep{1969ApL.....3..225R}, the proportion of pulsar radiation beams towards Earth, referred to as the beam fraction, depends on the radiation beam structure and the magnetic inclination angle. \citet{1994PhDT.......164G} introduced a relationship between the radius of the pulsar radiation beam and its rotational period: $\rho \propto P^{-1/2}$. Furthermore, \citeauthor{TM98} (\citeyear{TM98}; hereafter TM98) proposed an anti-correlation between the magnetic inclination angle $\alpha$ and the characteristic age $\tau=P/2\Dot{P}$. 

 Based on the evolutionary relationship between the magnetic inclination angle and characteristic age, the timescale for aligning the magnetic axis with the rotational axis is estimated to be approximately $10^7$ years. TM98 conducted an analysis of a substantial dataset of isolated pulsar polarization data. They proposed that the magnetic inclination-angle distribution of the entire pulsar population is not flat but highly concentrated at small angles. Additionally, the average beam fraction is only approximately 10\%. 
 Therefore, the estimated number of pulsars in the Galaxy is $(1.1\pm 0.2)\times 10^5$, and the birth rate of neutron stars is about $1.1\pm 0.2$ per century. Applying the TM98 beaming correction, the estimated surface density of pulsars near Earth is $470\pm50$ pulsars $\mathrm{kpc^{-2}}$, while the surface density of pulsars near the GC is $150\pm 80 $ pulsars $\mathrm{kpc^{-2}}$. It can also be estimated that the birth rate of neutron stars in the vicinity of Sun is approximately 4.7 $\pm$ 0.5 $\mathrm{kpc^{-2}\ Myr^{-1}}$.


\section{Conclusion}

Based on pulsars in the latest version of ATNF Pulsar Catalogue (v1.70) that were discovered in the Parkes Multibeam pulsar surveys, we described the radial distribution of pulsars in the Galaxy. Our results are summarized as follows:

  (i) The pulsar surface density increases inward from the Sun toward the Galactic Center to the point where the spiral arms begin, reaching its maximum value at a galactocentric radius of 3.91 $\pm$ 0.13 kpc.

  (ii) The surface density of pulsars around the Sun is 47 $\pm$ 8 and 470 $\pm$ 50 pulsars $\mathrm{kpc^{-2}}$, before and after the TM98 beaming correction, respectively. The birth rate of neutron stars in the vicinity of Sun is estimated to be approximately 4.7 $\pm$ 0.5 $\mathrm{kpc^{-2}\ Myr^{-1}}$.

  (iii) The pulsar surface density around the GC region ($R \sim 0.83$ kpc) is 18 $\pm$ 8 and 150 $\pm$ 80 pulsars $\mathrm{kpc^{-2}}$ before and after applying beaming correction, respectively. Based on the radial distribution function and Galactic structural parameters \citep{2014A&A...569A.125H}, the proportion of pulsars in the six structural components of the Galaxy were calculated, as shown in Table \ref{tab:spiralpara}.

  (iv) The total number of detectable pulsars in the Galaxy is (1.1 $\pm$ 0.2) $\times$ $10^{4}$ and (1.1 $\pm$ 0.2) $\times$ $10^{5}$ before and after the beaming correction, respectively. The birth rate of observable pulsars is estimated to be approximately 0.11 $\pm$ 0.02 per century, and after applying beaming correction, the birth rate of pulsar is found to be about 1.1 $\pm$ 0.2 per century.

The QiTai radio Telescope (QTT) in north-western China \citep{2023SCPMA..6689512W}, with its 110-meter aperture, has approximately three times the sensitivity of the Parkes radio telescope under the assumption of identical observation parameters and receiver temperature. The FAST radio telescope in south-eastern China \citep{nlj+11}, has an effective aperture of about 300 meters, and achieves a sensitivity approximately 22 times greater than that of Parkes. We can expect the accuracy of these numbers to improve as more sensitive pulsar surveys using these radio telescopes increase the available sample of pulsars.

\section*{Acknowledgement}
This work is supported the Major Science and Technology Program of Xinjiang Uygur Autonomous Region (No.2022A03013-4), the Zhejiang Provincial Natural
Science Foundation of China (No. LY23A030001), the National SKA Program of China (No.2020SKA0120100), the Natural Science Foundation of Xinjiiang Uygur Autonomous Region (No.2022D01D85), National Natural Science Foundation of China (No.12041304).

\bibliography{Pulsar_Radial}{}
\bibliographystyle{aasjournal}

\end{document}